\documentclass[conference]{IEEEtran}
\IEEEoverridecommandlockouts

\usepackage{tabularx}
\usepackage{cite}
\usepackage[dvips]{graphicx}
\usepackage[table, dvipsnames]{xcolor}
\usepackage{color, colortbl}
\usepackage{bbm}
\usepackage{url}
\definecolor{lavenderblue}{rgb}{0.8, 0.8, 1.0}
\definecolor{skyblue}{rgb}{0.53, 0.81, 0.92}
\definecolor{blizzardblue}{rgb}{0.67, 0.9, 0.93}
	\definecolor{aqua}{rgb}{0.0, 1.0, 1.0}
   \graphicspath{{../eps/}}
   \DeclareGraphicsExtensions{.eps}

\usepackage{mathrsfs}
\usepackage{amssymb}
\usepackage{yfonts}

\usepackage[cmex10]{amsmath}
\usepackage{algorithmicx}
\usepackage{algpseudocode}

\usepackage{algorithm}
\usepackage{graphicx}
\usepackage{array}
\usepackage{color}
\usepackage[caption=false,font=footnotesize]{subfig}
\usepackage{lettrine}
\usepackage{tikz}
\usepackage{verbatim}
\usepackage[bookmarks=false]{hyperref}

{}
\floatname{algorithm}{\textbf{Algorithm}}{}
\ifCLASSINFOpdf

\else

\fi
\begin{document}

\title{A Conditional Variational Framework for Channel Prediction in High-Mobility $6$G OTFS Networks\\
\thanks{This work was supported by Institute for Industrial Information Technology (inIT), Technische Hochschule Ostwestfalen-Lippe (TH-OWL), $32657$ Lemgo, Germany.}
}

\author{\IEEEauthorblockN{1\textsuperscript{st} Mohsen Kazemian}
\IEEEauthorblockA{\textit{institute industrial IT (inIT)} \\
\textit{technische
hochschule OWL (TH-OWL)}\\
Lemgo, Germany \\
mohsen.kazemian@th-owl.de}

\and
\IEEEauthorblockN{2\textsuperscript{nd} Jürgen Jasperneite}
\IEEEauthorblockA{\textit{institute industrial IT (inIT)} \\
\textit{technische
hochschule OWL (TH-OWL)}\\
Lemgo, Germany\\
juergen.jasperneite@th-owl.de}
}


\maketitle

\begin{abstract}
This paper proposes a machine learning (ML)–based method for channel prediction in high-mobility orthogonal time frequency space (OTFS) channels. In these scenarios, rapid variations caused by Doppler spread and time-varying multipath propagation lead to fast channel decorrelation, making conventional pilot-based channel estimation methods prone to outdated channel state information (CSI) and excessive overhead. Therefore, reliable channel prediction methods become essential to support robust detection and decoding in OTFS systems. In this paper, we propose conditional variational autoencoder for channel prediction (CVAE$4$CP) method, which learns the conditional distribution of OTFS delay–Doppler channel coefficients given physical system and mobility parameters. By incorporating these parameters as conditioning information, the proposed method enables the prediction of future channel coefficients before their actual realization, while accounting for inherent channel uncertainty through a low-dimensional latent representation. The proposed framework is evaluated through extensive simulations under high-mobility conditions. Numerical results demonstrate that CVAE$4$CP consistently outperforms a competing learning-based baseline in terms of normalized mean squared error (NMSE), particularly at high Doppler frequencies and extended prediction horizons. These results confirm the effectiveness and robustness of the proposed approach for channel prediction in rapidly time-varying OTFS systems.
\end{abstract}

\begin{IEEEkeywords}
channel prediction, CSI, OTFS, conditional variational autoencoder, $6$G.
\end{IEEEkeywords}

\IEEEpeerreviewmaketitle

\section{Introduction}
Orthogonal time frequency space (OTFS) modulation has been widely recognized as a strong waveform candidate for sixth-generation ($6$G) wireless networks, particularly in high-mobility systems such as unmanned aerial vehicles (UAVs), non-geostationary orbit (NGSO) satellite links, and high-speed autonomous transportation, where wireless channels exhibit rapid time variations \cite{2otfsr}. In such environments, conventional orthogonal frequency-division multiplexing (OFDM) suffers from severe intercarrier interference (ICI) due to large Doppler spreads, leading to performance degradation and reduced spectral efficiency  \cite{2khodspr}, \cite{2arxive}. OTFS mitigates these limitations by multiplexing information symbols in the delay–Doppler (DD) domain, effectively exploiting full time–frequency diversity and improving robustness against Doppler-induced impairments, even in uncoded systems. However, the rapidly time-varying nature of high-mobility channels makes accurate channel estimation and prediction a critical challenge in OTFS systems, as conventional pilot-based methods incur excessive overhead, leading to reduced spectral efficiency, and offer limited tracking capability in rapidly time-varying channels.

In high-mobility scenarios, rapid channel variations within the estimation interval often render the acquired channel state information (CSI) outdated before it can be effectively used. This limitation motivates the use of channel prediction techniques, which aim to infer future channel behavior from past observations, thereby reducing pilot overhead and improving spectral efficiency, while alleviating the burden of frequent channel estimation. However, the majority of existing channel prediction studies, such as \cite{2llmcp}, have primarily focused on OFDM-based systems, while only a limited number of works (e.g. \cite{2khod}) have addressed channel prediction in OTFS-based communications.

By contrast, extensive research has been devoted to OTFS channel estimation, which is commonly approached from two broad perspectives. One class of methods consists of physics-driven, model-based techniques that exploit parametric channel representations, typically assuming either limited Doppler-shift channels (LDSC) or continuous Doppler-spread channels (CDSC) \cite{22}. In this category, classical estimation techniques, including sparse Bayesian learning \cite{2sparse}, \cite{2bayesian} and compressed sensing methods \cite{2compressed}, have demonstrated effectiveness under the LDSC assumption but often encounter performance degradation when extended to CDSC scenarios due to increased channel complexity and higher pilot requirements. While model-based approaches benefit from interpretability and analytical structure, their performance often degrades in highly dynamic environments where channel evolution is nonlinear and difficult to capture with fixed parametric models.
The second category consists of data-driven or AI-based methods, which aim to learn channel behavior directly from data \cite{2aidriven}, \cite{2deeplearn}. AI-based methods are capable of learning complex channel behaviors directly from data without requiring explicit assumptions on channel statistics or temporal evolution. The work in  \cite{2aidriven} proposes a three-phase deep learning–assisted channel prediction framework for high-mobility MIMO-OTFS systems, where an autoencoder is used for channel feature compression, an RNN provides coarse channel prediction, and significant channel coefficients are reconstructed via least-squares estimation. This method relies on explicit detection and tracking of significant path indices and a multi-stage reconstruction pipeline, making it sensitive to path index prediction errors and limiting its ability to capture the full stochastic distribution of time-varying OTFS channels. However, by leveraging data-driven representations, these methods can adapt to diverse propagation conditions and capture intricate time-varying effects that are challenging to model analytically, making them particularly attractive for channel prediction in high-mobility OTFS systems.


Consequently, despite extensive progress in OTFS channel estimation, channel prediction in rapidly time-varying systems remains an open research problem, particularly when accounting for pilot overhead, stale channel estimates, and the limited ability of model-based methods to capture nonlinear channel evolution and stochastic multipath dynamics. These challenges motivate us to investigate an alternative channel prediction frameworks with a probabilistic and data-driven formulation, such as conditional variational learning, which can model the full conditional distribution of OTFS channel coefficients and effectively account for channel uncertainty under diverse mobility conditions.

In high-mobility propagation environments, OTFS channels exhibit pronounced randomness due to time-varying multipath propagation and Doppler effects. Consequently, even under the same known system configuration, such as Doppler spread, delay spread, antenna setup, and signal-to-noise ratio (SNR), multiple channel realizations may arise, highlighting the inherently stochastic nature of the underlying channel process. Unlike deterministic learning methods that yield a single point estimate, a conditional variational autoencoder (CVAE) provides a probabilistic framework for modeling the conditional distribution of channel coefficients given known system and propagation parameters \cite{2vae}. CVAEs are deep latent-variable generative models that extend the variational autoencoder (VAE) framework by incorporating auxiliary information into both the inference and generative networks. By conditioning the latent representation on side information, CVAEs enable scenario-controlled channel generation while preserving a compact stochastic latent space that captures uncertainty and variability beyond what can be explained by the observed conditioning variables alone. This capability makes CVAEs particularly well-suited for robust channel prediction in rapidly varying OTFS environments, where multiple plausible channel realizations must be accounted for. In this paper, we propose conditional variational autoencoder for channel prediction (CVAE$4$CP) method to model and predict the time-varying OTFS channel coefficients conditioned on key system and propagation parameters. The proposed model learns the conditional distribution of the delay–Doppler channel representation, where latent variables capture unobserved environmental factors driving channel variability. To improve modeling flexibility beyond conventional Gaussian assumptions, normalizing flows are incorporated into the latent space.
The main contributions of this paper are summarized as follows:

$\bullet$ We formulate the OTFS channel prediction problem as a conditional generative modeling task, where the distribution of delay–Doppler channel coefficients is learned given physical system parameters. This formulation explicitly accounts for the stochastic nature of time-varying OTFS channels, which is not captured by deterministic prediction approaches in prior studies.

$\bullet$ We develop a CVAE-based framework that leverages latent variables to model unobserved environmental and propagation effects influencing OTFS channel evolution. Unlike existing model-based or deterministic learning methods, the proposed approach captures multiple plausible channel realizations under the same operating conditions, enabling robust prediction in high-mobility scenarios.

$\bullet$ We incorporate normalizing flows into the CVAE latent space to overcome the limitations of Gaussian latent assumptions commonly adopted in prior learning-based channel models. This enhancement allows the proposed framework to represent complex, non-Gaussian delay–Doppler channel statistics and nonlinear temporal variations that are overlooked in existing OTFS channel prediction methods.


\section{System Model}


\begin{figure}[t]\label{fig:1}
\centering
        \includegraphics[width=3.4in]{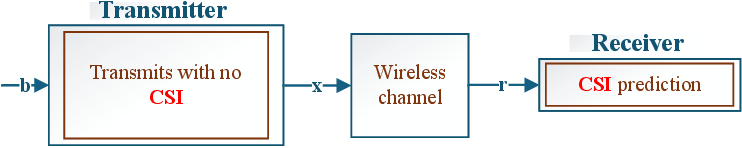}
     \caption{Wireless communication system with receiver-side CSI prediction.}
    \label{fig:1}
\end{figure}

OTFS is typically defined into two main structures: OFDM-based (OB)-OTFS and direct DD-domain (D$4$)-OTFS \cite{2khod}. In the first structure, the DD data symbols are transmitted to the channel in the time domain after a DD-to-time transformation using an inverse symplectic finite Fourier transform (ISFFT) block followed by a Heisenberg transform. In contrast, in D$4$-OTFS, the symbol data are transmitted directly in the DD domain without the intermediate step of transforming into the time domain. In this paper, we focus on D$4$-OTFS due to its lower structural complexity compared to OB-OTFS. Let $\mathbf{B}\in \mathbb{C}^{M\times N}$ in the DD domain and vector $\mathbf{x}=vec\left(\mathbf{B}\right)$ with length $M\times N$ be the data matrix and the transmitted signal, respectively. As shown in Fig. \ref{fig:1}, in high-mobility OTFS systems, CSI is typically predicted and utilized at the receiver for data detection, while the transmitter operates without instantaneous CSI and transmits data independently of the channel realization. The received signal at the receiver is defined as $r[{i}]=\sum_{l=0}^{L-1}
h[{i},l]x[{i}-l]+W[{i}]$, where $h[{i},l]$ is the wide-sense stationary uncorrelated scattering (WSSUS) Rayleigh fading coefficient corresponding to delay tap $l$ at index ${i}$, with ${i}\in \{0,...,MN-1\}$ representing the linearized index of the $M\times N$ DD-grid bins, and $W[{i}] \thicksim \mathcal{C}\mathcal{N}({0},{{{\sigma^2} }})$ denotes additive white Gaussian noise (AWGN).

\section{Proposed Method}\label{bb33}

In this section, we present the proposed CVAE$4$CP framework for receiver-side OTFS channel prediction. The framework leverages known system and propagation parameters as conditioning information and employs an expressive latent-variable model to capture the stochastic and time-varying nature of OTFS channels.

\subsection{Problem Formulation}\label{bb34}
Let $\mathbf{e}$ denote the vector of known system and propagation parameters, including Doppler spread, delay spread, SNR, antenna configuration, and OTFS grid dimensions. For learning purposes, we define
$\mathbf{X} = \left[ Re\{h\}^{T}, Im\{h\}^{T} \right]^{T} \in \mathbb{R}^{2D}$ containing the real and imaginary parts of the channel coefficient $h$. Given historical channel observations and their associated conditioning variables, the objective is to learn the conditional distribution $p(\mathbf{X}\mid\mathbf{e})$, which characterizes the stochastic relationship between physical channel parameters and OTFS channel realizations. This formulation naturally accounts for the fact that multiple channel realizations may arise under the same system configuration.

\subsection{Proposed CVAE-Based OTFS Channel Modeling}\label{bb34}
We employ a CVAE to model the conditional distribution $p(\mathbf{X} \mid \mathbf{e})$. The encoder network parameterized by $\phi$ approximates the conditional posterior $q_{\phi}({z} \mid \mathbf{X}, \mathbf{e})$, where ${z}\in\mathcal{Z}$ is a latent variable capturing unobserved environmental and propagation factors and $\mathcal{Z}$ denotes a low-dimensional latent space. The conditional decoder parameterized by $\theta$ models the likelihood $p_{\theta}(\mathbf{X} \mid {z}, \mathbf{e})$ and produces a reconstructed channel representation $\mathbf{\hat{X}}$. 

During training, the CVAE parameters are optimized using the following equation
\begin{multline}\label{eq11}
  \underset{\theta,\phi}{\mathrm{max}}~  \mathcal{L}(\theta, \phi; \mathbf{X}, \mathbf{e}) = \mathbb{E}_{q_{\phi}(z \mid \mathbf{X}, \mathbf{e})}\big[\log p_{\theta}(\mathbf{X} \mid z, \mathbf{e})\big] -\\ \mathcal{D}_{KL}\big(q_{\phi}(z \mid \mathbf{X}, \mathbf{e}) \| p_{\theta}(z \mid \mathbf{e})\big),
\end{multline}
where $\mathcal{L}(\theta, \phi; \mathbf{X}, \mathbf{e})$ denotes the conditional evidence lower bound
(ELBO), $\mathbb{E}_{q_{\phi}}[\cdot]$ represents the expectation with respect to the variational posterior, $\log p_{\theta}(\mathbf{X} \mid z, \mathbf{e})$ is the conditional log-likelihood term and $\mathcal{D}_{KL}$ denotes the 
Kullback-Leibler divergence measuring the discrepancy between the variational posterior $q_{\phi}(z \mid \mathbf{X}, \mathbf{e})$ and the conditional prior $p_{\theta}(z \mid \mathbf{e})$.

At inference time, given a new operating condition $\check{\mathbf{e}}$, channel realizations are generated by sampling $\mathbf{z} \sim p_{\theta}(z \mid \check{\mathbf{e}})$, and decoding via $\mathbf{\hat{X}}= g_\theta ({z}, \check{\mathbf{e}})$, where $g_\theta(\cdot)$ is the deterministic mapping implemented by the decoder network. This enables prediction of future OTFS channel coefficients without requiring instantaneous CSI at the transmitter.

\subsection{Latent Space Enhancement via Normalizing Flows}\label{bb34}
While conventional CVAE implementations typically assume a Gaussian latent distribution, such an assumption may be insufficient for modeling the complex, non-Gaussian statistics of time-varying OTFS channels, particularly under high-mobility conditions. To address this limitation, we enhance the expressiveness of the latent space by incorporating normalizing flows into the proposed framework. Specifically, an initial latent variable ${z}_0 \sim \mathcal{N}(\mathbf{0},\mathbf{I})       $ is transformed through a sequence of $K$ invertible and differentiable mappings such that ${z}_k=f_k({z}_{k-1})$, with $k\in \{1,...,K\}$, where each transformation $f_k(\cdot)$ is parameterized by a neural network and conditioned on $\mathbf{e}$. The resulting latent variable ${z}_k$ follows a more flexible distribution capable of approximating complex, multimodal channel statistics. Therefore, the conditional log-likelihood term is written as follows and replaces the standard Gaussian prior in the KL-divergence term of Eq. (\ref{eq11}):
\begin{equation}
\log p_{\theta}({z}_k \mid \mathbf{e})
=
\log p({z}_0)
-
\sum_{k=1}^{K}
\log
\Bigg|
\det
\left(
\frac{\partial f_k({z}_{k-1})}
{\partial {z}_{k-1}}
\right)
\Bigg|,
\end{equation}
therefore, the CVAE parameters are optimized using
\begin{multline}
  \underset{\theta,\phi}{\mathrm{max}}~  \mathcal{L}_{nf}(\theta, \phi; \mathbf{X}, \mathbf{e}) = \mathbb{E}_{q_{\phi}(z_0 \mid \mathbf{X}, \mathbf{e})}\big[\log p_{\theta}(\mathbf{X} \mid {z}_k, \mathbf{e})\big] -\\ \mathcal{D}_{KL}\big(q_{\phi}({z}_0 \mid \mathbf{X}, \mathbf{e}) \| p_{\theta}({z}_k \mid \mathbf{e})\big).
\end{multline}
This enhancement allows the proposed CVAE to capture nonlinear temporal variations and intricate delay–Doppler structures that are difficult to model using Gaussian latent assumptions. We summarize the design steps of our proposed method as the pseudo-code in Algorithm $1$.

\begin{algorithm}[t]\label{algo1}
\caption{The proposed CVAE$4$CP framework in OTFS systems.}
\textbf{Input:} Training pairs $\mathbf{X}_i$ and $\mathbf{e}_i$, epochs $E$ and batch size $\mathcal{B}$,
\begin{algorithmic}[1] 

\item \textbf{Training: for} each epoch {\textbf{do}}

\State{Compute $q_{\phi}({z} \mid \mathbf{X}, \mathbf{e})$}, $p_{\theta}(\mathbf{X} \mid {z}, \mathbf{e})$,

\State{Update $\phi$ and  $\theta$ by maximizing ELBO},
\item \textbf{end for}
\item \textbf{Inference:} Given a new condition $\check{\mathbf{e}}$,

\State{Sample $\mathbf{z} \sim p_{\theta}(z \mid \check{\mathbf{e}})$,}

\end{algorithmic}
\textbf{Output:} predicted OTFS channel coefficients $\mathbf{\hat{X}}= g_\theta ({z}, \check{\mathbf{e}})$.
\end{algorithm}

\section{Simulation Results}\label{bb33}
This section evaluates the performance of the proposed CVAE-based OTFS channel prediction framework through extensive Monte Carlo simulations. A synthetic dataset of $1000$ independent samples is constructed, where each sample pairs a $20$-dimensional vector $\mathbf{e}$ of physical channel and system parameters with the corresponding OTFS delay–Doppler channel coefficients under high-mobility operating conditions. Each channel realization is modeled in the delay–Doppler domain on an OTFS grid with $M=32$, $N=32$ and $L=6$ delay taps, yielding a complex-valued channel vector of length $32\times 32\times 6=6144$. By separating real and imaginary components, this representation is mapped to a $12288$-dimensional real-valued feature vector, which is concatenated with a $20$-dimensional vector of physical conditioning parameters. These parameters include the SNR randomly sampled between $0$ and $30$ dB, user speed ranging from $1$ to $60$ m/s, carrier frequency selected from $\{2.6,3.5,28\}$ GHz, bandwidth uniformly drawn between $5$ and $40$ MHz, subcarrier spacing chosen from $\{15,30,60\}$ kHz, cyclic prefix length selected from $\{0,16,32\}$, and MIMO transmit and receive antenna configurations independently selected from $\{1,2,4\}$. The resulting $12308$-dimensional input is projected by the CVAE encoder into a latent space with dimension $\mathbf{z}=48$. The dataset is randomly split into $N_{tr}=800$ training samples and $N_{te}=200$ testing samples, and the model is trained for $E=50$ epochs using the Adam optimizer with a learning rate $\alpha=10^{-3}$
and a batch size $\mathcal{B}=16$. Let $
NMSE = \frac{1}{N_{te}}
\sum_{i=1}^{N_{te}}
\frac{\left\| \mathbf{h}_i - \hat{\mathbf{h}}_i \right\|_2^2}
{\left\| \mathbf{h}_i \right\|_2^2}$. We evaluate the performance of our method on the test samples in terms of NMSE under varying Doppler frequencies and prediction horizons to assess the model's ability to capture temporal and mobility-induced OTFS channel dynamics.

Fig. $2$ demonstrates the NMSE performance versus the maximum Doppler frequency $f_D$ for OTFS channel prediction, comparing the proposed CVAE$4$CP framework with the RNN-based method in \cite{2aidriven}. As expected, NMSE increases for both approaches as the Doppler frequency rises due to the reduced temporal correlation of the channel under high-mobility conditions. Nevertheless, the proposed CVAE$4$CP method consistently outperforms the baseline across the entire Doppler range. In particular, at $f_D=5$ kHz CVAE$4$CP achieves an NMSE on the order of approximately $4\times 10^{-3}$, whereas the method in \cite{2aidriven} exhibits an NMSE of approximately $5\times 10^{-2}$, corresponding to more than one order of magnitude improvement. This result highlights the robustness of the proposed probabilistic framework in capturing mobility-induced channel variations in high-Doppler OTFS environments.

The NMSE as a function of the prediction horizon $\Delta$ is illustrated in Fig. $3$, where $\Delta$ denotes the number of OTFS frames ahead being predicted. As the prediction horizon increases, the NMSE of both methods degrades due to the increasing difficulty of long-term channel extrapolation. However, CVAE$4$CP maintains significantly lower prediction error compared to \cite{2aidriven} for all considered horizons. For instance, at $\Delta=10$, the proposed method achieves an NMSE of approximately $2\times 10^{-3}$, while the baseline error increases to the order of $2\times 10^{-1}$. This pronounced performance gap demonstrates the superior temporal generalization capability of CVAE$4$CP and its effectiveness in predicting future OTFS channel states over extended horizons.
\begin{figure}[t]
\centering
        \includegraphics[width=3.5in]{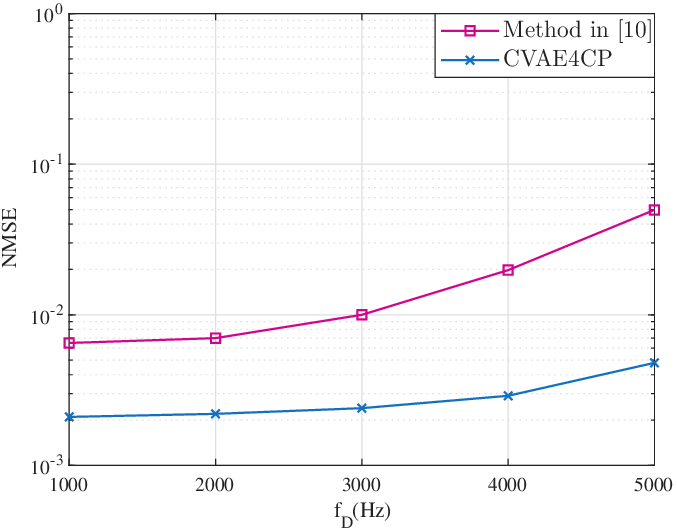}
     \caption{NMSE versus maximum Doppler frequency for OTFS channel prediction, comparing the proposed CVAE4CP with the RNN-based method in \cite{2aidriven}.}
    \label{fig:4}
\end{figure}

\begin{figure}[t]
\centering
        \includegraphics[width=3.5in]{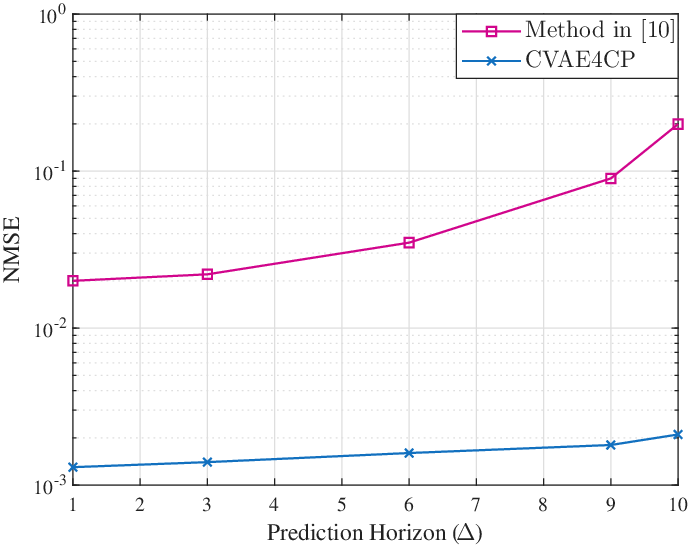}
     \caption{NMSE versus prediction horizon (number of OTFS frames ahead), comparing the proposed CVAE$4$CP with the RNN-based method in \cite{2aidriven}.}
    \label{fig:4}
\end{figure}

\section{Conclusion}\label{defin6}
In this paper, we have proposed a ML–based framework for channel prediction in high-mobility OTFS systems, thereby ensuring timely channel awareness, while reducing pilot overhead and improving spectral utilization. First, a CVAE was developed to model the conditional distribution of delay–Doppler channel coefficients given physical system and mobility parameters, enabling the prediction of future channel coefficients prior to their realization. Second, the latent space representation was enhanced through probabilistic modeling, allowing the framework to capture inherent channel uncertainty and variability beyond deterministic point estimation. Simulation results demonstrate that the proposed CVAE-based predictor effectively tracks the temporal and mobility-induced dynamics of OTFS channels. Compared with a baseline method, the proposed approach achieves significantly lower NMSE, particularly under high Doppler frequencies and extended prediction horizons. These results confirm the robustness and effectiveness of the proposed framework for channel prediction in rapidly time-varying OTFS environments. Future work will investigate extensions to multi-user OTFS systems and online adaptation under non-stationary channel conditions.

\leavevmode%


\ifCLASSOPTIONcaptionsoff
  \newpage
\fi
\bibliographystyle{IEEEtran}
\bibliography{keylatex}
\end{document}